\documentclass[3p]{elsarticle}

\usepackage{lineno,hyperref}
\usepackage{amsmath}
\modulolinenumbers[5]

\journal{Journal of Computational Physics}

\begin{document}

\begin{frontmatter}

\title{Structure-preserving operators for thermal-nonequilibrium hydrodynamics}


\author[mymainaddress]{Takashi Shiroto\corref{mycorrespondingauthor}}
\ead{tshiroto@rhd.mech.tohoku.ac.jp}
\cortext[mycorrespondingauthor]{Corresponding author}

\author[mymainaddress]{Soshi Kawai}

\author[mymainaddress]{Naofumi Ohnishi}

\address[mymainaddress]{Department of Aerospace Engineering, Tohoku University, 6-6-01 Aramaki-Aza-Aoba, Aoba-ku, Sendai, Miyagi 980-8579, Japan}

\begin{abstract}
Radiation hydrodynamics simulations based on the one-fluid two-temperature model may violate the law of energy conservation
because the governing equations are expressed in a nonconservative formulation.
Here, we maintain the important physical requirements by employing a strategy based on the key concept that the mathematical structures
associated with the conservative and nonconservative equations are preserved, even at the discrete level.
To this end, we discretize the conservation laws and transform them via exact algebraic operations.
The proposed scheme maintains the global conservation errors within the round-off level.
In addition, a numerical experiment concerning the shock tube problem suggests that the proposed scheme well agrees with
the jump conditions at the discontinuities regulated by the Rankine--Hugoniot relationship.
The generalized derivation allows us to employ arbitrary central difference, artificial dissipation, and Runge--Kutta methods.
\end{abstract}


\end{frontmatter}

\linenumbers

\section{Introduction}\label{sec:1}
Radiation hydrodynamics (RHD)\cite{MihalasMihalas} is one of the major techniques employed in laser-plasma simulations.
In RHD, a neutral charge is assumed for a fluid composed of ions and electrons. 
This assumption allows large-scale plasma simulations to be performed, as the grid interval is not limited
by the Debye length, unlike the conventional particle-in-cell (PIC) method.
This type of simulation has been employed in laser-plasma simulations that model,
for example, the implosion dynamics of inertial confinement fusion (ICF)\cite{Nuckolls1972}.

One of the most simplified RHD simulations combines radiative transfer and one-temperature hydrodynamics,
which are employed by the FastRad3D code\cite{Bates2016}.
However, thermal nonequilibrium typically exists between the ions and electrons of a laser plasma.
The incident laser is absorbed by the inverse bremsstrahlung process, and the energy is deposited on the electrons.
Thus, the one-fluid two-temperature (1F2T) model has been employed in order to investigate plasma hydrodynamics
more accurately \cite{Fujioka2004-2,Dangelo2013,Tanaka2015}.
The governing equations of the 1F2T model excluding viscous and heat conduction effects are expressed in the form
\begin{align}
\frac{\partial \rho}{\partial t}+\nabla\cdot(\rho \mathbf{u})=0,\label{eq:1.1}\\
\frac{\partial \rho \mathbf{u}}{\partial t}+\nabla\cdot(\rho \mathbf{uu}+p_\mathrm{i}+p_\mathrm{e})=0,\label{eq:1.2}\\
\frac{\partial \rho e_\mathrm{i}}{\partial t}+\nabla\cdot(\rho e_\mathrm{i}\mathbf{u})+p_\mathrm{i}\nabla\cdot\mathbf{u}=0,\label{eq:1.3}\\
\frac{\partial \rho e_\mathrm{e}}{\partial t}+\nabla\cdot(\rho e_\mathrm{e}\mathbf{u})+p_\mathrm{e}\nabla\cdot\mathbf{u}=0,\label{eq:1.4}
\end{align}
where $\rho$ is the mass density, $\mathbf{u}=^\mathrm{T}[u,v,w]$ is the flow velocity,
$e$ is the specific internal energy, and $p$ is the pressure.
The subscripts ``$\mathrm{i}$'' and ``$\mathrm{e}$'' denote the ions and electrons, respectively.
Note that nonhydrodynamic energy transports such as radiative transfer are neglected.
These nonconservative equations have been used in many RHD simulations \cite{Colombier2005,Cao2015}.

Three main methods of solving the 1F2T model exist.
The first (hereafter referred to as ``nonconservative method") discretizes the governing equations directly
with the Lagrangian or arbitrary Lagrangian--Eulerian (ALE) grids\cite{Hirt1974}.
This method has been adopted for use in the RHD codes known as HYDRA\cite{Marinak2001}, DRACO\cite{Keller1999,Radha2005}, and
PINOCO\cite{Nagatomo2007}, which are specialized for ICF implosions.
However, because the nonconservative equations are discretized directly using the artificial dissipation,
the law of energy conservation is not satisfied, especially near the discontinuities such as a shock wave.
For example, it has been reported that Helios-CR\cite{MacFarlane2006}, a one-dimensional RHD code, has an error of several percent
with regard to the law of energy conservation in the typical cases.
The second discretization technique is based on approximate Riemann solvers and methods of this kind are
employed by CRASH\cite{vanderHolst2011}, FLASH\cite{Fryxell2000}, RAGE\cite{Gittings2008},
and RAICHO\cite{Ohnishi2012}, which are primarily implemented in astrophysical simulations.
Note that some of these codes incorporate adaptive mesh refinement (AMR) for the performance of high-resolution simulations,
and cross-code comparisons are also conducted\cite{Joggerst2014}.
These codes are designed to solve the Euler equation and the hydrodynamic heating is
artificially divided between the ions and electrons in proportion to the pressure ratio.
Therefore, the results are not guaranteed to converge to the exact solutions,
because the governing equations are not discretized directly.
However, the energy conservation is satisfied even in the discrete form.

The most recently developed method (hereafter, ``conservative method") is implemented in the ASTER\cite{Igmenshchev2016} code
developed at the Laboratory of Laser Energetics, University of Rochester.
This code solves the conservation laws of mass, momentum, and total energy,
along with the energy equation given in Eq.~(\ref{eq:1.4}).
Thus, the conservation law of total energy is satisfied automatically
and the numerical solutions converge to the exact solutions, provided the solutions do not possess any discontinuities.
However, relatively high-intensity lasers ($\sim$10$^{15}\mathrm{\ W/cm^2}$) are usually employed in typical
laser-plasma investigations\cite{Smalyuk2010,Hu2012,Haines2016,Shiroto2016}, which yield strong shock waves.
The internal energies of the ions and electrons may not be accurately determined near these shock waves.
This problem has a critical impact on RHD simulations, because the emission and absorption coefficients
of radiative transfer are strongly dependent on the electron internal energy.

In this paper, we propose a structure-preserving scheme that can overcome the above problems.
The proposed scheme should achieve the following requirements:
(i) solve the partially nonconservative equations (\ref{eq:1.1})--(\ref{eq:1.4}),
(ii) maintain the error of energy conservation at the round-off level, and
(iii) get a good agreement with exact solutions including discontinuities.
In Section~\ref{sec:2}, we derive the conservative scheme through a discussion using the discrete mathematics.
The mathematical keys for the exact derivation are the product and quotient rules in the discrete calculus; therefore,
these tools are also introduced.
The numerical methods employed in this study are described in Section~\ref{sec:3}.
Furthermore, in Section~\ref{sec:4}, the results of a numerical experiment on an extended shock-tube problem
are discussed, and the superiority of the proposed scheme over the existing
nonconservative and conservative methods is demonstrated.
The accuracy of the proposed scheme is verified in Section~\ref{sec:5}
via a linear advection problem concerning entropy waves.
Section~\ref{sec:6} presents the conclusion of this article.

\section{Mathematical background of the proposed approach}\label{sec:2}

\subsection{Causes of energy conservation law violation}\label{sec:2.0}
Equations~(\ref{eq:1.1})--(\ref{eq:1.4}) can be mathematically transformed into the law of energy conservation by
using the product and quotient rules,
\begin{align}
\frac{\partial \left(\rho e_\mathrm{i}+\rho e_\mathrm{e}+\frac12\rho |\mathbf{u}|^2\right)}{\partial t}+
\nabla\cdot\left(\rho e_\mathrm{i}\mathbf{u}+\rho e_\mathrm{e}\mathbf{u}+\frac12\rho |\mathbf{u}|^2\mathbf{u}
+p_\mathrm{i}\mathbf{u}+p_\mathrm{e}\mathbf{u}\right)=0.\label{eq:1.5}
\end{align}
Physically, the law of energy conservation should be maintained by solving Eqs.~(\ref{eq:1.1})--(\ref{eq:1.4}).
However, when these equations are solved in discrete form, the energy conservation is
often violated. Sometimes, this error can be fatal to the simulations.
The conservation law is violated because the energy equations (\ref{eq:1.3}) and (\ref{eq:1.4})
are expressed in nonconservative formulation.
To explain why the conservation laws are violated in the discrete form,
we convert Eq.~(\ref{eq:1.1}) into nonconservative form using the product rule, as an example. Thus, Eq.~(\ref{eq:1.1}) becomes
\begin{align}
\frac{\partial \rho}{\partial t}+\rho\nabla\cdot\mathbf{u}+\mathbf{u}\cdot\nabla\rho=0.\label{eq:1.6}
\end{align}
The forward-time central-space method is then applied to Eqs.~(\ref{eq:1.1}) and (\ref{eq:1.6}) to obtain
the discretized equations in the finite difference method (FDM). Thus,
\begin{align}
\frac{\rho^{n+1}_j-\rho^n_j}{\Delta t}+\frac{\langle\rho u\rangle^n_{j^+}
-\langle \rho u\rangle^n_{j^-}}{\Delta x}=0,\label{eq:1.7}\\
\frac{\rho^{n+1}_j-\rho^n_j}{\Delta t}+\rho^n_j\frac{\langle u\rangle^n_{j^+}-\langle u\rangle^n_{j^-}}{\Delta x}
+u^n_j\frac{\langle \rho\rangle^n_{j^+}-\langle \rho\rangle^n_{j^-}}{\Delta x}=0,\label{eq:1.8}
\end{align}
where ``$\langle \rangle$'' denotes an arbitrary interpolation operator at the half points ($j^+=j+1/2, j^-=j-1/2$),
and $n$ and $j$ are the indices of time and space, respectively.
Note that the spatial derivatives in the $y$ and $z$ directions are omitted for simplicity.
Of course, Eqs.~(\ref{eq:1.7}) and (\ref{eq:1.8}) correspond to Eqs.~(\ref{eq:1.1}) and (\ref{eq:1.6}), respectively.
We take the summation of Eqs.~(\ref{eq:1.7}) and (\ref{eq:1.8}) over the computational domain in order to discuss the global conservation in the discrete system; thus, we obtain
\begin{align}
\sum_{j=1}^N\frac{\rho^{n+1}_j-\rho^n_j}{\Delta t}+\sum_{j=1}^N\frac{\langle \rho u\rangle^n_{j^+}
-\langle\rho u\rangle^n_{j^-}}{\Delta x}=0,\label{eq:1.9}\\
\sum_{j=1}^N\frac{\rho^{n+1}_j-\rho^n_j}{\Delta t}+\sum_{j=1}^N\rho^n_j\frac{\langle u\rangle^n_{j^+}-\langle u\rangle^n_{j^-}}{\Delta x}
+\sum_{j=1}^Nu^n_j\frac{\langle\rho\rangle^n_{j^+}-\langle\rho\rangle^n_{j^-}}{\Delta x}=0,\label{eq:1.10}
\end{align}
where $N$ is the number of grids.
The advection terms of Eq.~(\ref{eq:1.9}) are cancelled, yielding
\begin{align}
\sum_{j=1}^N\frac{\rho^{n+1}_j-\rho^n_j}{\Delta t}+\frac{\langle \rho u\rangle^n_{N+1/2}
-\langle\rho u\rangle^n_{1/2}}{\Delta x}=0.\label{eq:1.11}
\end{align}
Therefore, the total mass of the discrete system does not change
if the periodic [$\langle \rho u\rangle_{1/2}=\langle\rho u\rangle_{N+1/2}$]
or Neumann [$\langle\rho u\rangle_{1/2}=\langle\rho u\rangle_{N+1/2}=0$] boundaries are applied.
In contrast, Eq.~(\ref{eq:1.10}) violates the conservation laws, as the advection terms are not generally cancelled.

The above example is the most simple explanation of why the conservation laws are violated in the discrete form,
and of why the mathematical structure of the product rule should be preserved, even at the discrete level.
Below, we discuss our idea of how to construct a structure-preserving scheme.

\subsection{Product and quotient rules in discrete form}\label{sec:2.1}
The nonconservative energy equation is obtained from the conservation laws of mass, momentum, and energy.
Although the derivation is performed using the product and quotient rules,
these formulae may violate the global conservation in the discrete form.
Therefore, we introduce the discrete product and quotient rules.

First, we introduce the product rule in discrete form:
\begin{align}
\frac{f^{n+1}g^{n+1}-f^ng^n}{\Delta t}&=\frac{f^{n+1}g^{n+1}-f^{n+1}g^n+f^{n+1}g^n-f^ng^n}{\Delta t},\notag\\
&=f^{n+1}\frac{g^{n+1}-g^n}{\Delta t}+\frac{f^{n+1}-f^n}{\Delta t}g^n,\label{eq:2.2}
\end{align}
where $f$ and $g$ are arbitrary functions depending on $t$.
Algebraic operations performed on the numerator, which add and subtract $f^{n+1}g^n$,
correspond to the same approach as that used to prove the original formula in the differential form.
The other forms of the product rule are obtained in a similar manner, where
\begin{align}
\frac{f^{n+1}g^{n+1}-f^ng^n}{\Delta t}=\frac{f^{n+1}+f^n}{2}\frac{g^{n+1}-g^n}{\Delta t}+
\frac{f^{n+1}-f^n}{\Delta t}\frac{g^{n+1}+g^n}{2}.\label{eq:2.4}
\end{align}

The quotient rule in discrete form is also obtained via the same strategy, yielding
\begin{align}
\frac{f^{n+1}/g^{n+1}-f^n/g^n}{\Delta t}&=\frac{g^{n+1}+g^n}{2g^{n+1}g^n}\frac{f^{n+1}-f^n}{\Delta t}-
\frac{f^{n+1}+f^n}{2g^{n+1}g^n}\frac{g^{n+1}-g^n}{\Delta t}.\label{eq:2.10}
\end{align}
The quotient rule of Eq.~(\ref{eq:2.10}) is mathematically equivalent to the product rule of Eq.~(\ref{eq:2.4});
however, we must expand the kinetic energy term as a function of the mass and momentum.
The quotient rule is useful when a conservative variable appears in the denominator of a function.

\subsection{Derivation of structure-preserving scheme}\label{sec:2.2}
In this subsection, we derive a scheme which exactly satisfies
the law of energy conservation with solving the internal energy of ions and electrons.
It is the structure-preserving scheme that the mathematical structure associating
the conservative formulation with the nonconservative one is strictly maintained.
The Euler equation is discretized using the FDM approach, such that
\begin{align}
\frac{\rho^{n+1}_j-\rho^n_j}{\Delta t}+\frac{\langle\rho u\rangle^n_{j^+}
-\langle\rho u\rangle^n_{j^-}}{\Delta x}=0,\label{eq:2.11}\\
\frac{(\rho u)^{n+1}_j-(\rho u)^n_j}{\Delta t}+
\frac{\langle\rho u^2+p\rangle^n_{j^+}-
\langle\rho u^2+p\rangle^n_{j^-}}{\Delta x}=0,\label{eq:2.12}\\
\frac{(\rho e+\frac12\rho u^2)^{n+1}_j-(\rho e+\frac12\rho u^2)^n_j}{\Delta t}
+\frac{\langle\rho eu+\frac12\rho u^3+p u\rangle^n_{j^+}-
\langle\rho eu+\frac12\rho u^3+pu\rangle^n_{j^-}}{\Delta x}=0,\label{eq:2.13}
\end{align}
where $p=p(\rho, e)$ is the total pressure.
As shown in Section~\ref{sec:2.0}, the conservation laws of mass, momentum, and energy are strictly satisfied in the system.
Furthermore, although we employ the Euler explicit method here,
the scheme can be extended to the Runge--Kutta (RK) methods.
Proof of this is given in \ref{sec:a}.
The multidimensional description is given in \ref{sec:b}.

The structure-preserving scheme is obtained by expanding Eq.~(\ref{eq:2.13}).
The time derivative of the kinetic energy is expanded in the form
\begin{align}
\frac{(\rho u^2)^{n+1}_j-(\rho u^2)^n_j}{\Delta t}&=
\frac{1}{\Delta t}\left[ \left\{\frac{(\rho u)^2}{\rho}\right\}^{n+1}_j - \left\{\frac{(\rho u)^2}{\rho}\right\}^n_j\right],\notag\\
&=\frac{\rho^{n+1}_j+\rho^n_j}{2\rho^{n+1}_j\rho^n_j}\frac{\{(\rho u)^2\}^{n+1}_j-\{(\rho u)^2\}^n_j}{\Delta t}-
\frac{\{(\rho u)^2\}^{n+1}_j+\{(\rho u)^2\}^n_j}{2\rho^{n+1}_j\rho^n_j}\frac{\rho^{n+1}_j-\rho^n_j}{\Delta t},\notag\\
&=\frac{\rho^{n+1}_j+\rho^n_j}{2\rho^{n+1}_j\rho^n_j}\left\{(\rho u)^{n+1}_j+(\rho u)^n_j\right\}
\frac{(\rho u)^{n+1}_j-(\rho u)^n_j}{\Delta t}
-\frac{\{(\rho u)^2\}^{n+1}_j+\{(\rho u)^2\}^n_j}{2\rho^{n+1}_j\rho^n_j}\frac{\rho^{n+1}_j-\rho^n_j}{\Delta t},\notag\\
&=\frac{\rho^{n+1}_j+\rho^n_j}{2\rho^{n+1}_j\rho^n_j}\left\{(\rho u)^{n+1}_j+(\rho u)^n_j\right\}
\left(-\frac{\langle \rho u^2+p\rangle^n_{j^+}-
\langle\rho u^2+p\rangle^n_{j^-}}{\Delta x}\right)\notag\\
&+\frac{\{(\rho u)^2\}^{n+1}_j+\{(\rho u)^2\}^n_j}{2\rho^{n+1}_j\rho^n_j}
\frac{\langle\rho u\rangle^n_{j^+}-\langle\rho u\rangle^n_{j^-}}{\Delta x}.\label{eq:2.18}
\end{align}
Here, the quotient and product rules of Eqs.~(\ref{eq:2.10}) and (\ref{eq:2.4}), respectively, are used
in the derivation.
Some readers may feel that this expansion is unnecessary, as both
$\rho^{n+1}_j$ and $u^{n+1}_j$ have already been obtained in Eqs.~(\ref{eq:2.11}) and (\ref{eq:2.12}).
However, this operation is necessary as it clarifies the contributions of the ion and electron pressures on
the right hand side (RHS) of Eq.~(\ref{eq:2.18}).
This clarification is required in order to prove that the proposed scheme qualifies as the discretized equations
of Eqs.~(\ref{eq:1.3}) and (\ref{eq:1.4}).
In order to obtain the nonconservative formulation,
Eq.~(\ref{eq:2.18}) is substituted into Eq.~(\ref{eq:2.13}), yielding
\begin{align}
\frac{(\rho e)^{n+1}_j-(\rho e)^n_j}{\Delta t}+
\frac{\langle\rho eu\rangle^n_{j^+}-\langle\rho eu\rangle^n_{j^-}}{\Delta x}+
\frac{\langle pu\rangle^n_{j^+}-\langle pu\rangle^n_{j^-}}{\Delta x}
-\frac{\rho^{n+1}_j+\rho^n_j}{4\rho^{n+1}_j \rho^n_j}\{(\rho u)^{n+1}_j+(\rho u)^n_j\}
\frac{\langle p\rangle^n_{j^+}-\langle p\rangle^n_{j^-}}{\Delta x}=\notag\\
-\frac{(\rho^2 u^2)^{n+1}_j+(\rho^2 u^2)^n_j}{4\rho^{n+1}_j\rho^n_j}
\frac{\langle \rho u\rangle^n_{j^+}-\langle \rho u\rangle^n_{j^-}}{\Delta x}
+\frac{\rho^{n+1}_j+\rho^n_j}{4\rho^{n+1}_j \rho^n_j}\{(\rho u)^{n+1}_j+(\rho u)^n_j\}
\frac{\langle \rho u^2\rangle^n_{j^+}-\langle \rho u^2\rangle^n_{j^-}}{\Delta x}\notag\\
-\frac12\frac{\langle \rho u^3\rangle^n_{j^+}-\langle \rho u^3\rangle^n_{j^-}}{\Delta x}.\label{eq:2.19}
\end{align}
This equation provides important information regarding the energy conservation in the 1F2T model.
The left hand side (LHS) of Eq.~(\ref{eq:2.19}) is a discretized formulation of the
nonconservative energy equation
\begin{align}
\frac{\partial (\rho e)}{\partial t}+
\frac{\partial (\rho e u)}{\partial x}+
p\frac{\partial u}{\partial x}=0
\qquad \mathrm{or} \qquad
\frac{\partial (\rho e)}{\partial t}+
\frac{\partial(\rho e u)}{\partial x}+
\frac{\partial (p u)}{\partial x}-u\frac{\partial p}{\partial x}=0.\label{eq:2.21}
\end{align}
The rest of the terms on the RHS are error terms that are cancelled out when $\Delta t \to 0$ and $\Delta x \to 0$.
The information on this error is lost in Eq.~(\ref{eq:2.21}); therefore,
it is difficult to reconstruct the error terms using intuitive discretizations.
Furthermore, note that Eq.~(\ref{eq:2.21}) is the sum of Eqs.~(\ref{eq:1.3}) and (\ref{eq:1.4}),
which correspond to the first law of thermodynamics for ions and electrons.
This relationship must be satisfied even at the discrete level; therefore,
Eq.~(\ref{eq:2.19}) provides a constraint condition that regulates the law of energy conservation in the discrete 1F2T model.

Finally, the structure-preserving scheme for the 1F2T model can be written in the form
\begin{align}
\frac{(\rho e_\mathrm{s})^{n+1}_j-(\rho e_\mathrm{s})^n_j}{\Delta t}+
\frac{\langle\rho e_\mathrm{s}u\rangle^n_{j^+}-\langle\rho e_\mathrm{s}u\rangle^n_{j^-}}{\Delta x}+
\frac{\langle p_\mathrm{s}u\rangle^n_{j^+}-\langle p_\mathrm{s}u\rangle^n_{j^-}}{\Delta x}
-\frac{\rho^{n+1}_j+\rho^n_j}{4\rho^{n+1}_j \rho^n_j}\{(\rho u)^{n+1}_j+(\rho u)^n_j\}
\frac{\langle p_\mathrm{s}\rangle^n_{j^+}-\langle p_\mathrm{s}\rangle^n_{j^-}}{\Delta x}\notag\\
=
-\frac{(\rho^2 u^2)^{n+1}_j+(\rho^2 u^2)^n_j}{8\rho^{n+1}_j\rho^n_j}
\frac{\langle \rho u\rangle^n_{j^+}-\langle \rho u\rangle^n_{j^-}}{\Delta x}
+\frac{\rho^{n+1}_j+\rho^n_j}{8\rho^{n+1}_j \rho^n_j}\{(\rho u)^{n+1}_j+(\rho u)^n_j\}
\frac{\langle \rho u^2\rangle^n_{j^+}-\langle \rho u^2\rangle^n_{j^-}}{\Delta x}\notag\\
-\frac14\frac{\langle \rho u^3\rangle^n_{j^+}-\langle \rho u^3\rangle^n_{j^-}}{\Delta x}.\label{eq:2.22}
\end{align}
The terms incorporating the internal energy ($e$) or pressure ($p$) in Eq.~(\ref{eq:2.19}) can be easily separated
because of the physical requirements.
However, the other terms have no explicit restrictions for partition.
Equation~(\ref{eq:2.22}) distributes the error terms equally among the ions and electrons; hence, the symmetry of Eqs.~(\ref{eq:1.3}) and (\ref{eq:1.4}) is retained.
The Rankine--Hugoniot relationship and the law of equipartition can only be reproduced using this approach.
This will be mentioned in the later verification.
Note that the extension to the multitemperature model is straightforward, because of the law of equipartition.

\subsection{Shock capturing method}\label{sec:2.2.1}
When solving the hydrodynamic field with the discontinuities,
the shock capturing method is required to obtain the entropy solutions.
Here, we derive the structure-preserving scheme including artificial dissipation terms.
The discretized Euler equation Eqs.~(\ref{eq:2.11})--(\ref{eq:2.13}) is modified as
\begin{align}
\frac{\rho^{n+1}_j-\rho^n_j}{\Delta t}+\frac{\langle\rho u\rangle^n_{j^+}
-\langle\rho u\rangle^n_{j^-}}{\Delta x}=0,\label{eq:2.23}\\
\frac{(\rho u)^{n+1}_j-(\rho u)^n_j}{\Delta t}+
\frac{\langle\rho u^2+p\rangle^n_{j^+}-
\langle\rho u^2+p\rangle^n_{j^-}}{\Delta x}=
\frac{\langle A\rangle^n_{j^+}-\langle A\rangle^n_{j^-}}{\Delta x},\label{eq:2.24}\\
\frac{(\rho e+\frac12\rho u^2)^{n+1}_j-(\rho e+\frac12\rho u^2)^n_j}{\Delta t}
+\frac{\langle\rho eu+\frac12\rho u^3+p u\rangle^n_{j^+}-
\langle\rho eu+\frac12\rho u^3+pu\rangle^n_{j^-}}{\Delta x}=
\frac{\langle B\rangle^n_{j^+}-\langle B\rangle^n_{j^-}}{\Delta x},\label{eq:2.25}
\end{align}
where $A$ and $B$ are the artificial dissipations required to capture the discontinuities.
The nonconservative equations about the internal energy of ions and electrons
are derived by the same way in Sec.~\ref{sec:2.2}:
\begin{align}
\frac{(\rho e_\mathrm{s})^{n+1}_j-(\rho e_\mathrm{s})^n_j}{\Delta t}+
\frac{\langle\rho e_\mathrm{s}u\rangle^n_{j^+}-\langle\rho e_\mathrm{s}u\rangle^n_{j^-}}{\Delta x}+
\frac{\langle p_\mathrm{s}u\rangle^n_{j^+}-\langle p_\mathrm{s}u\rangle^n_{j^-}}{\Delta x}
-\frac{\rho^{n+1}_j+\rho^n_j}{4\rho^{n+1}_j \rho^n_j}\{(\rho u)^{n+1}_j+(\rho u)^n_j\}
\frac{\langle p_\mathrm{s}\rangle^n_{j^+}-\langle p_\mathrm{s}\rangle^n_{j^-}}{\Delta x}\notag\\
=-\frac{\rho^{n+1}_j+\rho^n_j}{8\rho^{n+1}_j \rho^n_j}\{(\rho u)^{n+1}_j+(\rho u)^n_j\}
\frac{\langle A\rangle^n_{j^+}-\langle A\rangle^n_{j^-}}{\Delta x}
+\frac{\langle B_\mathrm{s}\rangle^n_{j^+}-\langle B_\mathrm{s}\rangle^n_{j^-}}{\Delta x}
-\frac{(\rho^2 u^2)^{n+1}_j+(\rho^2 u^2)^n_j}{8\rho^{n+1}_j\rho^n_j}
\frac{\langle \rho u\rangle^n_{j^+}-\langle \rho u\rangle^n_{j^-}}{\Delta x}\notag\\
+\frac{\rho^{n+1}_j+\rho^n_j}{8\rho^{n+1}_j \rho^n_j}\{(\rho u)^{n+1}_j+(\rho u)^n_j\}
\frac{\langle \rho u^2\rangle^n_{j^+}-\langle \rho u^2\rangle^n_{j^-}}{\Delta x}
-\frac14\frac{\langle \rho u^3\rangle^n_{j^+}-\langle \rho u^3\rangle^n_{j^-}}{\Delta x}.\label{eq:2.26}
\end{align}
Note that the first term on the RHS about $A$, which is associated with the viscosity,
is equally separated between ions and electrons since the viscosity transfers
the momentum whose contributions of ions and electrons are inseparable.

\subsection{Summary of the proposed approach}\label{sec:2.2.2}
The summary of the proposed approach toward construction of the structure-preserving scheme is as follows:

\begin{enumerate}
\item Discretize the conservation laws of mass, momentum, and total energy Eqs.~(\ref{eq:2.27})--(\ref{eq:2.29})
so that the conservation laws are automatically satisfied even in the discrete level.
\item Expand the time derivative of kinetic energy in Eq.~(\ref{eq:2.29}) using the discrete product and quotient rules
to obtain the nonconservative energy equation about $\rho (e_\mathrm{i}+e_\mathrm{e})$ with error terms.
\item Separate the contributions of ions and electrons in the discretized nonconservative energy equation.
The terms with the pressure $p_\mathrm{s}$ and the specific internal energy $e_\mathrm{s}$ are
separated by the physically accurate way.
The rest terms only including the density $\rho$ and the velocity $u$ are mathematically equally shared
by assuming the law of equipartition.
\item Solve the discretized equations about the density $\rho$, the momentum $\rho u$,
the internal energy of ions $\rho e_\mathrm{i}$, and that of electrons $\rho e_\mathrm{e}$.
\end{enumerate}

\begin{align}
\frac{\partial \rho}{\partial t}+\frac{\partial (\rho u)}{\partial x}=0,\label{eq:2.27}\\
\frac{\partial (\rho u)}{\partial t}+\frac{\partial (\rho u^2+p_\mathrm{i}+p_\mathrm{e})}{\partial x}=0,\label{eq:2.28}\\
\frac{\partial (\rho e_\mathrm{i}+\rho e_\mathrm{e}+\frac12\rho u^2)}{\partial t}+
\frac{\partial (\rho e_\mathrm{i}u+\rho e_\mathrm{e}u+\frac12\rho u^3+p_\mathrm{i}u+p_\mathrm{e}u)}{\partial x}=0.\label{eq:2.29}
\end{align}

The above methodology is also available if the artificial dissipation terms are
included for the purpose of the shock capturing method.

\section{Numerical implementations}\label{sec:3}
In this study, the fourth-order Pad\'{e}-type interpolation\cite{Kobayashi1999} is utilized, with
\begin{align}
\frac14 \langle f\rangle_{j-1/2}+\langle f\rangle_{j+1/2}+
\frac14 \langle f\rangle_{j+3/2}=\frac32(f_j+f_{j+1}),\label{eq:3.1}
\end{align}
where $f$ is the flux of an arbitrary conservative equation.
This interpolation operator has the linearity as follows:
\begin{align}
\langle f_1+f_2 \rangle = \langle f_1 \rangle + \langle f_2 \rangle,\label{eq:3.15}\\
\langle \alpha f_1 \rangle = \alpha \langle f_1 \rangle ,\label{eq:3.16}
\end{align}
where $\alpha$ is an arbitrary constant.
Time integration is performed using the third-order total variation diminishing (TVD) RK method\cite{shu1988}.

Artificial dissipations are modeled using the bulk viscosity $\beta$ and the thermal conductivity $\kappa_\mathrm{s}$
for ions and electrons, where
\begin{align}
A=\beta\frac{\partial u}{\partial x},\label{eq:3.4}\\
B_\mathrm{s}=\frac12u\beta\frac{\partial u}{\partial x}+
\kappa_\mathrm{s}\frac{\partial e_\mathrm{s}}{\partial x}.\label{eq:3.5}
\end{align}
These transport coefficients are modeled in a similar manner to the
localized artificial diffusivity (LAD) scheme\cite{Kawai2008,Kawai2010a}
\begin{align}
\beta=C_\beta \rho f_\mathrm{sw}
\left|\frac{\partial^r}{\partial x^r}\left(\frac{\partial u}{\partial x}\right)\right|\Delta x^{r+2},\label{eq:3.7}\\
\kappa_\mathrm{s}=C_\kappa \frac{\rho a}{e_\mathrm{i}+e_\mathrm{e}}
\left|\frac{\partial^r e_\mathrm{s}}{\partial x^r}\right|\Delta x^{r+1},\label{eq:3.8}\\
f_\mathrm{sw}=H\left(-\frac{\partial u}{\partial x}\right),\label{eq:3.9}
\end{align}
where $H$ is the Heaviside function, $a$ is the sound speed, and $r=4$, according to the typical LAD usage.
$C_\beta$ and $C_\kappa$ are the nondimensional parameters of the LAD scheme and
fixed to $5$ and $2$, respectively.
High-order derivatives obtained from these compact schemes have noisy profiles;
thus, the obtained $\beta$ and $\kappa_\mathrm{s}$ should be smeared using the appropriate truncated Gaussian blur\cite{Cook2004}.
The first and fourth derivatives are derived using the fourth-order compact schemes:
\begin{align}
\frac14\left(\frac{\partial f}{\partial x}\right)_{j-1}+\left(\frac{\partial f}{\partial x}\right)_{j}+
\frac14\left(\frac{\partial f}{\partial x}\right)_{j+1}=\frac32\frac{f_{j-1}-f_{j+1}}{2\Delta x},\label{eq:3.10}\\
\frac14\left(\frac{\partial^4 f}{\partial x^4}\right)_{j-1}+\left(\frac{\partial^4 f}{\partial x^4}\right)_{j}+
\frac14\left(\frac{\partial^4 f}{\partial x^4}\right)_{j+1}=\frac32\frac{f_{j-2}-
4f_{j-1}+6f_{j}-4f_{j+1}+f_{j+2}}{\Delta x^4}.\label{eq:3.11}
\end{align}
Note that tridiagonal matrices appearing in the compact schemes are solved using the Thomas algorithm.

Hydrodynamic simulations with compact differences are often coupled with low-pass filters\cite{Lele1992,Gaitonde2000},
owing to the stabilization of numerical dispersion especially at high wavenumbers.
Low-pass filters are usually applied to the conservative variables.
The 1F2T model only has three conservation laws of mass, momentum, and energy
although the number of governing equations Eqs.~(\ref{eq:1.1})--(\ref{eq:1.4}) is four.
Therefore, the filtering scheme should be applied to the 1F2T model carefully.
We derive a constraint condition that should be preserved through the filtering operations:
\begin{align}
\rho^\star=\lceil\rho\rfloor,\label{eq:3.12}\\
u^\star=\frac{\lceil\rho u\rfloor}{\lceil\rho\rfloor},\label{eq:3.13}\\
e^\star_\mathrm{i}=\frac{\lceil\rho e_\mathrm{i}\rfloor}{\lceil\rho\rfloor}+
\frac14\left(\frac{\lceil\rho u^2\rfloor}{\lceil\rho\rfloor}-
\frac{\lceil\rho u\rfloor^2}{\lceil\rho\rfloor^2}\right),\label{eq:3.14}\\
e^\star_\mathrm{e}=\frac{\lceil\rho e_\mathrm{e}\rfloor}{\lceil\rho\rfloor}+
\frac14\left(\frac{\lceil\rho u^2\rfloor}{\lceil\rho\rfloor}-
\frac{\lceil\rho u\rfloor^2}{\lceil\rho\rfloor^2}\right),\label{eq:3.15}
\end{align}
where the superscript ``$\star$'' denotes the filtered quantities
and ``$\lceil \ \rfloor$'' is the filtering operator.
The momentum and total energy composed of the filtered primitive variables are
\begin{align}
\rho^\star u^\star=\lceil\rho u\rfloor,\\
\rho^\star e^\star_\mathrm{i}+\rho^\star e^\star_\mathrm{e}+\frac12 \rho^\star (u^\star)^2=
\lceil \rho e_\mathrm{i}+\rho e_\mathrm{e}+\frac12\rho u^2 \rfloor,
\end{align}
where we assume that the filtering operator also has the linearity.
Therefore, the conservation laws are maintained when the filtering operators are used like
Eqs.~(\ref{eq:3.12})--(\ref{eq:3.15}).
In this study, an eighth-order compact filter\cite{Shiroto2017a} is used in the final step of the RK integration
to stabilize the numerical dispersion, such that
\begin{align}
\alpha_\mathrm{f} g_{j-1/2}+ g_{j+1/2}+ \alpha_\mathrm{f} g_{j+3/2}=
\sum_{n=0}^3 b_n(1-2\alpha_\mathrm{f})(f_{j+1+n}+f_{j-n}),\label{eq:3.2}\\
\lceil f \rfloor_j = f_j - (g_{j+\frac12}-g_{j-\frac12}),\label{eq:3.3}
\end{align}
where $b_0=-\frac{35}{256}$, $b_1=\frac{21}{256}$,
$b_2=-\frac{7}{256}$, and $b_3=\frac{1}{256}$.
The filtering parameter $\alpha_\mathrm{f}$ is set to 0.495.

\section{Verification via shock tube problem}\label{sec:4}
The shock tube problem is employed to verify the effects of the conservative and
nonconservative schemes, as this problem includes discontinuous solutions, i.e.,
the shock wave and contact discontinuity\cite{Sod1978}.
Here, we extend the shock tube problem to the 1F2T model, so that the proposed scheme can be verified
using the exact solution.

We assume that a diaphragm separating the high- and low-pressure sections of a tube bursts at $t=0$ and $x=0.5$.
The computational domain is $0 \le x \le 1$ and the number of grids is 201.
The initial conditions are almost identical to the original Sod's problem:
$\rho_\mathrm{L}=1$, $\rho_\mathrm{R}=0.125$, $u_\mathrm{L}=u_\mathrm{R}=0$,
$e_{\mathrm{i,L}}=1.5$, and $e_{\mathrm{i,R}}=e_{\mathrm{e,L}}=e_{\mathrm{e,L}}=1$,
where the subscripts ``$\mathrm{L}$'' and ``$\mathrm{R}$'' denote the areas on the left and right sides
of the diaphragm, respectively.
The quantity $e_\mathrm{i}+e_\mathrm{e}$ is identical to the original Sod's problem. Therefore,
these conditions simply add the temperature nonequilibrium between the ions and electrons to the original problem.
Note that $e$ is associated with $p$ by the
thermally and calorically ideal equations of state (EOS), with the ratio of specific heat $\gamma=1.4$.
The boundaries satisfy the Neumann condition such that the summations of the mass and energy
are fixed to the initial values.

Here we show the exact solution of the 1F2T shock tube problem.
The governing equations can be described by the following quasi-linear form:
\begin{align}
\begin{bmatrix}
\dfrac{\partial \rho}{\partial t} \\[2ex]
\dfrac{\partial u}{\partial t} \\[2ex]
\dfrac{\partial e_\mathrm{i}}{\partial t} \\[2ex]
\dfrac{\partial e_\mathrm{e}}{\partial t}
\end{bmatrix}
+
\begin{bmatrix}
u & \rho & 0 & 0 \\[2ex]
\dfrac{(\gamma-1)(e_\mathrm{i}+e_\mathrm{e})}{\rho} & u & \gamma-1 & \gamma-1 \\[2ex]
0 & (\gamma-1)e_\mathrm{i} & u & 0 \\[2ex]
0 & (\gamma-1)e_\mathrm{e} & 0 & u
\end{bmatrix}
\begin{bmatrix}
\dfrac{\partial \rho}{\partial x} \\[2ex]
\dfrac{\partial u}{\partial x} \\[2ex]
\dfrac{\partial e_\mathrm{i}}{\partial x} \\[2ex]
\dfrac{\partial e_\mathrm{e}}{\partial x}
\end{bmatrix}
=
\begin{bmatrix}
0 \\[2ex] 0 \\[2ex] 0 \\[2ex] 0
\end{bmatrix}.\label{eq:4.1}
\end{align}
This is a hyperbolic system which has the eigenvectors $\mathbf{k}$ and the corresponding eigenvalues $\lambda$:
\begin{align}
\mathbf{k}_1=\begin{bmatrix}
\rho \\ -a \\ (\gamma-1)e_\mathrm{i} \\ (\gamma-1)e_\mathrm{e}
\end{bmatrix},\quad
\mathbf{k}_2=\begin{bmatrix}
\rho \\ 0 \\ -e_\mathrm{i} \\ 0
\end{bmatrix},\quad
\mathbf{k}_3=\begin{bmatrix}
\rho \\ 0 \\ 0 \\ -e_\mathrm{e}
\end{bmatrix},\quad
\mathbf{k}_4=\begin{bmatrix}
\rho \\ a \\ (\gamma-1)e_\mathrm{i} \\ (\gamma-1)e_\mathrm{e}
\end{bmatrix},\label{eq:4.2}\\
\lambda_1 = u-a,\quad \lambda_2=\lambda_3=u,\quad \lambda_4=u+a,\label{eq:4.3}
\end{align}
where $a=\sqrt{\gamma(p_\mathrm{i}+p_\mathrm{e})/\rho}$ is the sound speed.
The first and fourth eigenmodes denote the pressure waves while
the second and third ones are the entropy waves for ions and electrons, respectively.
According to the eigenstructure, the ratio of internal energy $e_\mathrm{s}/(e_\mathrm{i}+e_\mathrm{e})$
only changes at the contact surface.
The exact solutions for each internal energies are easily derived from the original Sod's solution with this fact.

\begin{figure}
\includegraphics[width=0.8\textwidth]{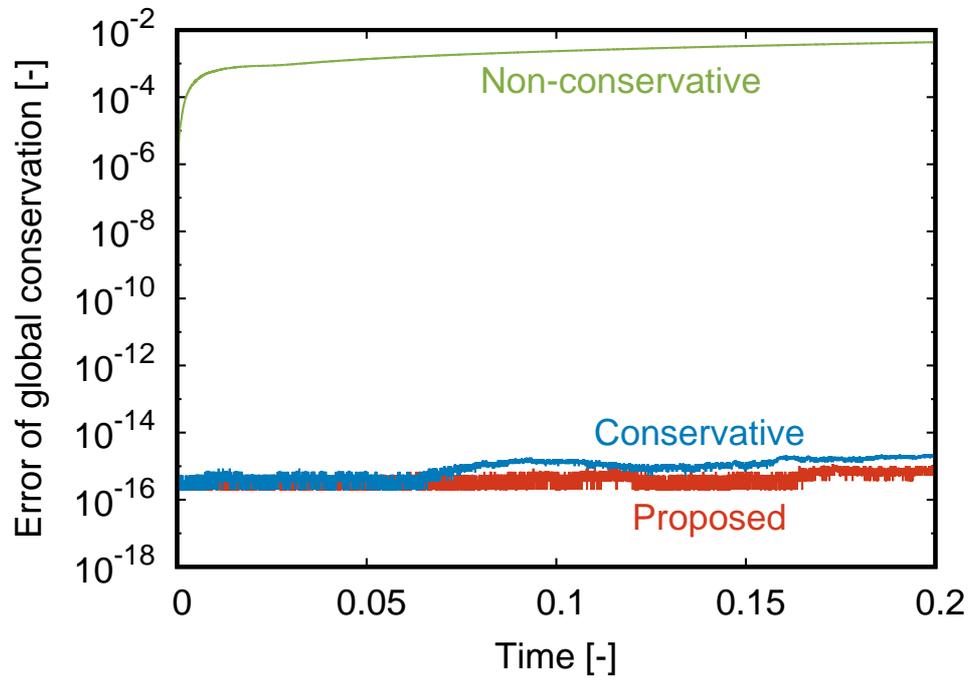}
\centering
\caption{\label{fig:4.2} (Color online) Time histories of errors with respect to law of energy conservation for examined schemes.
}
\end{figure}

Figure~\ref{fig:4.2} shows a time history of the errors with respect to the global conservation of energy.
As mentioned in Section~\ref{sec:1}, the nonconservative method obviously violates the law of energy conservation,
because this method discretizes the nonconservative equations.
On the other hand, the conservative and proposed methods keep the errors within the machine zero level
of the double-precision floating-point numbers ($\sim$2$~\times 10^{-16}$).
This is because the law of energy conservation is discretized directly using these techniques.
The behaviors of these errors differ from each other; however, this discrepancy has no physical meaning,
as these are the round-off rather than truncation errors.
Therefore, the conservative and proposed schemes are proven to be conservative.
Moreover, the round-off errors of these schemes do not linearly accumulate in the numerical experiments.
This is primarily because we employ the finite-volume-method (FVM) like
interpolation and filtering schemes\cite{Kobayashi1999,Shiroto2017a}.
Note that this is the key to maintain the global conservation errors at the round-off level.

\begin{figure}
\includegraphics[width=0.8\textwidth]{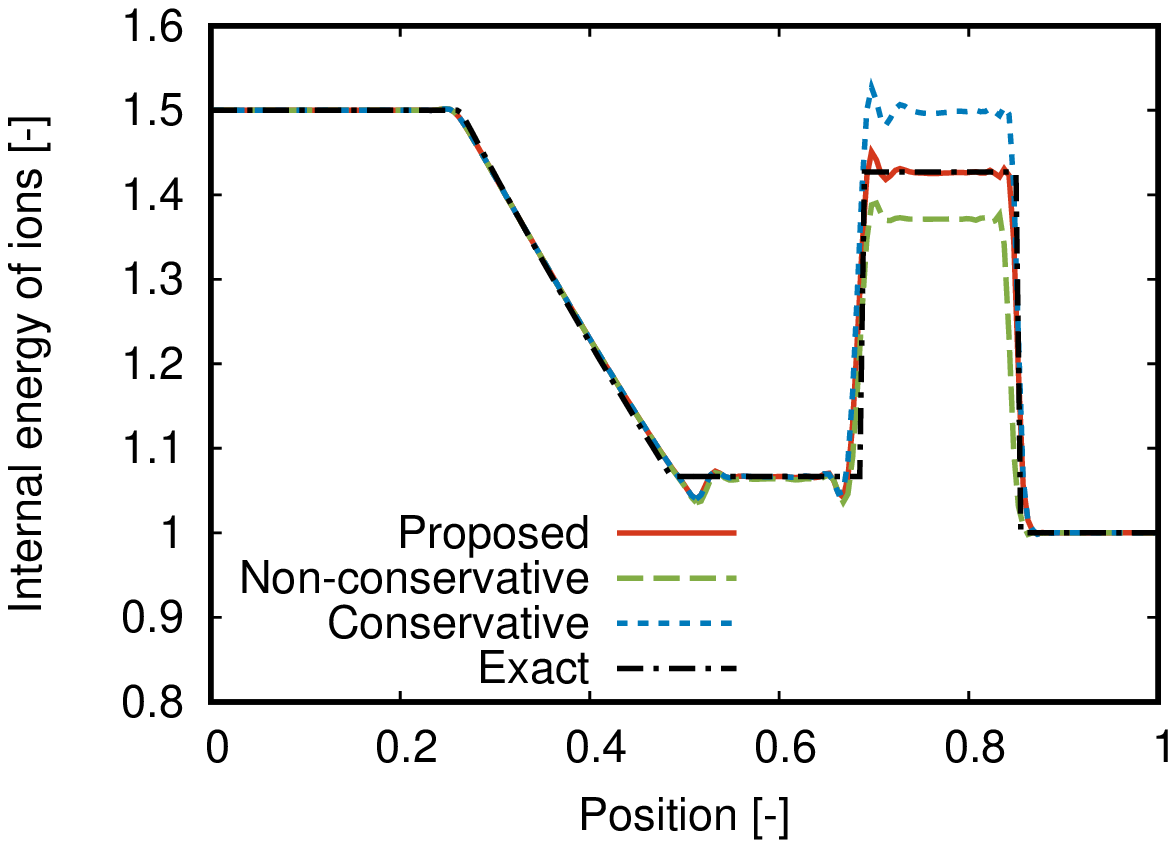}
\centering
\caption{\label{fig:4.3} (Color online) Internal energy profiles for ions $e_\mathrm{i}$ at $t=0.2$, for examined schemes and exact solution.
}
\end{figure}

\begin{figure}
\includegraphics[width=0.8\textwidth]{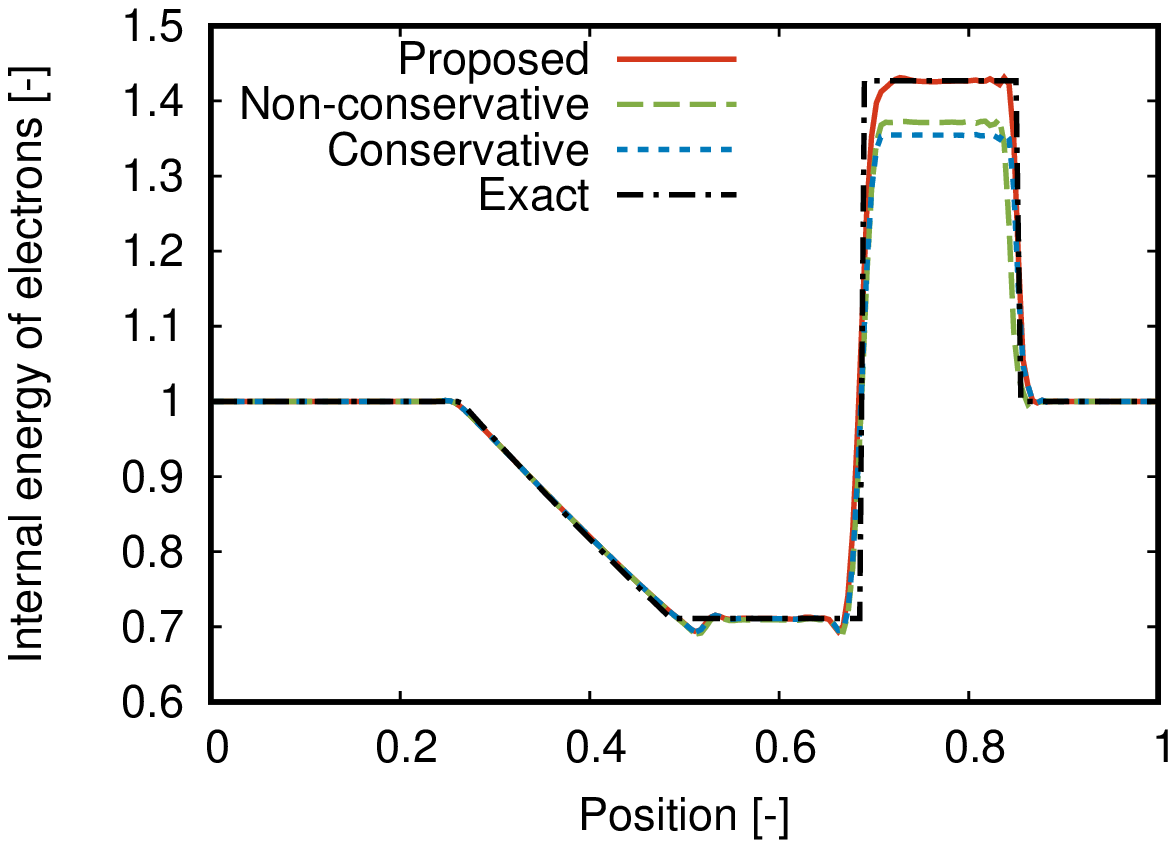}
\centering
\caption{\label{fig:4.4} (Color online) Internal energy profiles for electrons $e_\mathrm{e}$ at $t=0.2$, for examined schemes and exact solution.
}
\end{figure}

Figures~\ref{fig:4.3} and \ref{fig:4.4} are spatial profiles of $e_\mathrm{i}$ and $e_\mathrm{e}$, respectively.
The proposed scheme well agrees with the exact solutions including the discontinuities.
However, profiles obtained by the existing nonconservative and conservative methods clearly deviate from
the exact solutions between the shock wave and contact surface.
The exact solutions are associated with the Rankine--Hugoniot relationship;
thus, they are influenced by the local principles such as the conservation laws.
Therefore, Figures~\ref{fig:4.3} and \ref{fig:4.4} indicate that the nonconservative and conservative schemes
violate some of these principles.
For the nonconservative scheme, $e_\mathrm{i}$ and $e_\mathrm{e}$ remain identical
between the discontinuities, while $e_\mathrm{i}+e_\mathrm{e}$
is lower than the exact solution.
In other words, the nonconservative scheme maintains the law of equipartition but violates the law of energy conservation
at the discrete level.
In contrast, the conservative scheme reproduces the spatial profile of $e_\mathrm{i}+e_\mathrm{e}$
but cannot maintain the equilibrium of the ions and electrons between the discontinuities.
This finding suggests that the conservative scheme maintains the law of energy conservation,
but the law of equipartition is violated near the discontinuities.
Note that the law of equipartition is strongly related to the symmetry of Eqs.~(\ref{eq:1.3}) and (\ref{eq:1.4}).
Thus, this symmetry should be preserved at the discrete level.
Both the nonconservative scheme, which directly discretizes the nonconservative equations,
and the proposed scheme determine $e_\mathrm{i}$ and $e_\mathrm{e}$ with the symmetric formulation.
In contrast, although the energy equation of Eq.~(\ref{eq:1.4}) is discretized directly by the conservative scheme,
$e_\mathrm{i}$ is indirectly obtained from the other discretized equations;
this explains why the existing conservative scheme violates the law of equipartition.
The derivation of the proposed scheme is unique;
the mathematical structures of the governing equations must be preserved
in order to maintain the important physical principles.

\section{Accuracy verification}\label{sec:5}
\begin{figure}
\includegraphics[width=0.8\textwidth]{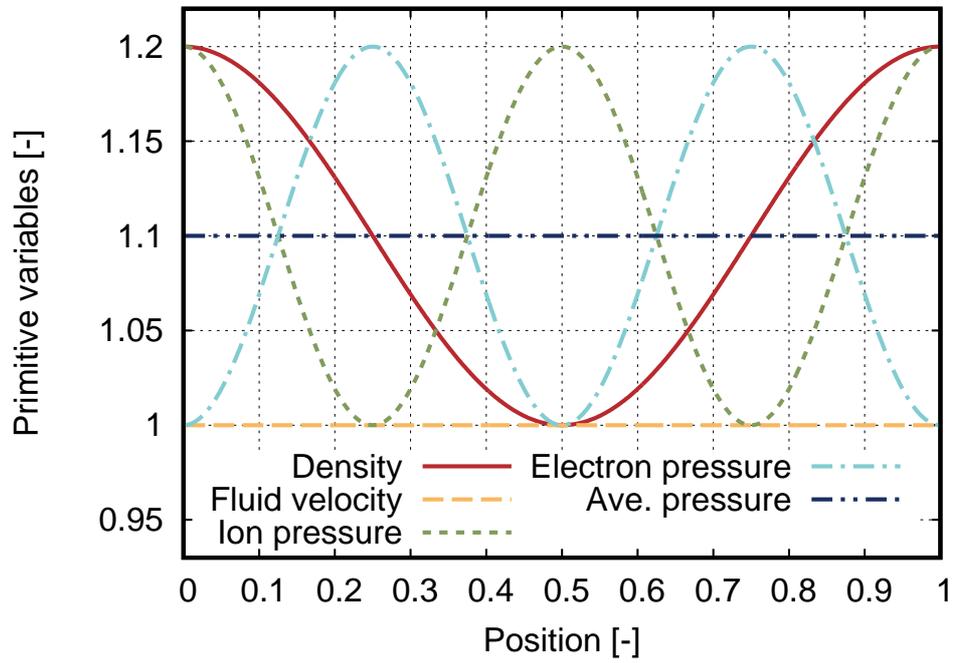}
\centering
\caption{\label{fig:5.1} (Color online) Initial conditions of linear advection problem.
}
\end{figure}

The spatial accuracy is assessed to check whether
the formal accuracy is reproduced by the proposed scheme.
The 1F2T hydrodynamic equations are decomposed into an eigenstructure comprised of two pressure waves
and the entropy waves of the ions and electrons.
We introduce this test problem to examine the spatial accuracy of the 1F2T model:
\begin{align}
\rho = 1.1+0.1\cos(2\pi x),\label{eq:5.1}\\
u = 1,\label{eq:5.2}\\
p_\mathrm{i} = 1.1+0.1\cos(4\pi x), \label{eq:5.3}\\
p_\mathrm{e} = 1.1-0.1\cos(4\pi x), \label{eq:5.4}\\
\gamma=1.4, \label{eq:5.5}
\end{align}
where $x\in[0,1]$. The initial profiles are depicted in Fig.~\ref{fig:5.1}.
Again, the EOS is thermally and calorically ideal.
These conditions eliminate the pressure waves which correspond to the nonlinear field.
Therefore, this is a linearized problem involving ion and electron entropy waves,
whose solutions are easily obtained via the hyperbolic solvers.
A periodic condition is applied to the boundaries, so that the initial conditions are recovered at $t=1$.
The time interval $\Delta t=10^{-5}$ is sufficiently small to make the temporal error negligible.

\begin{figure}
\includegraphics[width=0.8\textwidth]{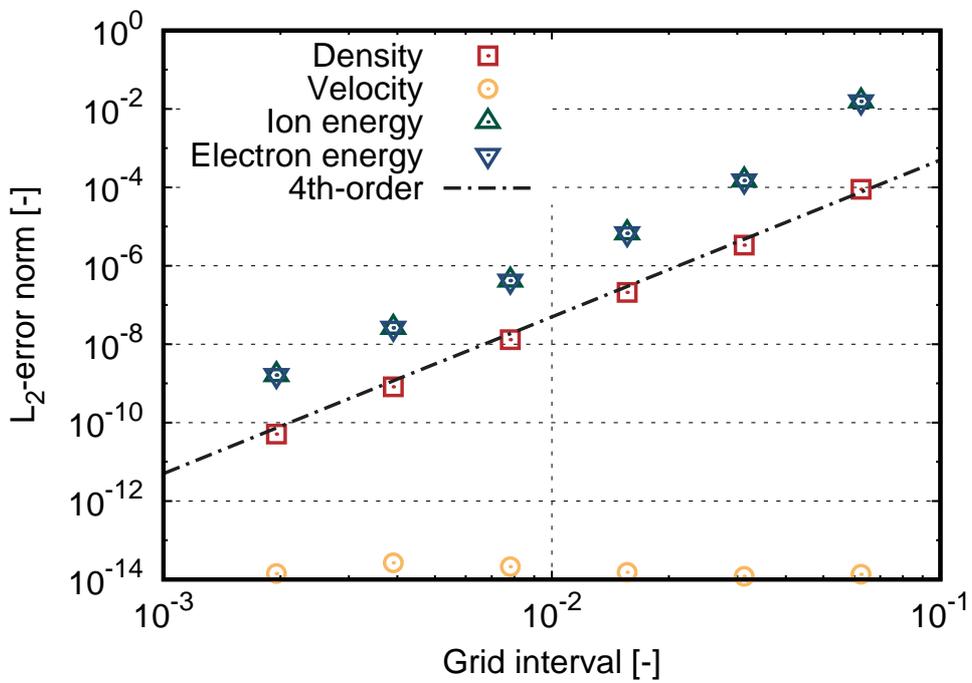}
\centering
\caption{\label{fig:5.3} (Color online) Accuracy verification for all primitive variables.
}
\end{figure}

Figure~\ref{fig:5.3} is a log-log graph to illustrate the accuracy verification.
The results show that $\rho$, $e_\mathrm{i}$, and $e_\mathrm{e}$
have the fourth-order formal accuracy of the compact scheme (Eq.~(\ref{eq:3.1})).
Note that $e_\mathrm{i}$ and $e_\mathrm{e}$ are not discretized separately;
the discretizations are only applied to the conservation laws of mass, momentum, and energy.
On the other hand, the spatial accuracy of $u$ cannot be observed via this numerical experiment,
because $u$ and the static pressure $p_\mathrm{i}+p_\mathrm{e}$ are initially constant in the entire region.
Equation~(\ref{eq:2.12}) is simplified by assuming $u^n_j=U=\mathrm{const.}$ and $p^n_j=\mathrm{const.}$, such that
\begin{align}
\rho^{n+1}\frac{u^{n+1}_j-u^n_j}{\Delta t}+U\frac{\rho^{n+1}_j-\rho^n_j}{\Delta t}
+U\frac{\langle \rho u\rangle^n_{j^+}-\langle \rho u\rangle^n_{j^-}}{\Delta x}=0.\label{eq:5.6}
\end{align}
Here, the relationship $\langle \rho u^2\rangle=U\langle \rho u\rangle$ is valid at the discrete level, because of
the linearity of the interpolation operators.
Hence, $u^{n+1}_j=u^n_j$ is obtained by substituting Eq.~(\ref{eq:2.11}) into Eq.~(\ref{eq:5.6}).
This is why the $u$ errors are independent of the
grid interval, so that the error-norm always maintains the round-off level.

Note that the discretizations are not based on the FVM but, rather, on the FDM,
although FVM-like schemes of interpolation and filtering are utilized in this investigation.

\section{Conclusions}\label{sec:6}
In this article, a structure-preserving scheme for the 1F2T hydrodynamic equations is
proposed, with the aim of improving the reliability of compressible RHD simulations.
The proposed scheme exactly satisfies the conservation laws
and thus the Rankine--Hugoniot relationship.

The key of constructing a physically accurate scheme that satisfies the important physical principles is
to maintain the mathematical structure of the governing equations, even in the discrete form.
Specifically, the product rule and the symmetry of the energy equations of each species must be maintained.
Therefore, the proposed approach does not discretize the energy equations of the ions and electrons directly
but, rather, discretizes the energy conservation law using the FDM approach.
The ion and electron energy equations in the discrete form are derived using the product rule;
this is the same strategy as that used to derive the nonconservative equations in differential form.
This derivation yields error terms in the energy equations,
which should be separated equally in accordance with the law of equipartition.

Verification via the shock tube problem demonstrates that the proposed scheme
maintains the global conservation error to within the round-off level and well agrees with
the Rankine--Hugoniot relationship of the 1F2T model.
In other words, the proposed scheme strictly preserves the conservation laws of
mass, momentum, and energy, and the law of equipartition in the discrete form.
Moreover, accuracy verification based on the linear advection of entropy waves reveals that
the proposed scheme yields the formal accuracy.

Although the proposed scheme possesses the prefer features explained above,
some issues remain toward practical RHD simulations, such as those considering the ICF implosion.
For example, the scheme must be generalized to curvilinear coordinates for
spherical\cite{Kidder1974}/cylindrical\cite{Piriz2002} implosions, non-ideal EOS,
and magnetohydrodynamics (MHD) for magnetized fast ignition \cite{Fujioka2013,Nagatomo2015}.
This problem may be solved by our strategy using a previous work about MHD scheme \cite{Kawai2013}.
Furthermore, our approach may be used in hypersonic hydrodynamics of re-entries.
It is modeled by multitemperature hydrodynamics regarding the translational, rotational,
vibrational and excitation modes \cite{Park1989,Sakai2001}.

\section*{Acknowledgments}
This work was supported by a Grant-in-Aid from the Japan Society for the Promotion of Science (JSPS) Fellows, No. 15J02622.
T.S. wishes to thank Dr. Atsushi Sunahara (Purdue University) for valuable discussions
on the physical background of the 1F2T model.

\appendix
\section{Usage of Runge--Kutta method in proposed scheme}\label{sec:a}
High-order time integration is important for the performance of high-resolution simulations.
Here, the implementations of the RK methods in the proposed scheme are presented.

\subsection{First-order RK method}\label{sec:a1}
The first-order RK method is identical to the Euler explicit method.
The terms in Eq.~(\ref{eq:2.22}) can be classified into the following three components:
\begin{align}
ADV^{n+1,n,n}_{\mathrm{s},j}=DFS^{n+1,n,n}_{\mathrm{s},j}+ERR^{n+1,n,n}_{\mathrm{s},j},\label{eq:a.1}\\
ADV^{k,l,m}_{\mathrm{s},j}\equiv \frac{(\rho e_\mathrm{s})^k_j-(\rho e_\mathrm{s})^l_j}{\Delta t}+
\frac{\langle \rho e_\mathrm{s}u \rangle^m_{j^+}-\langle \rho e_\mathrm{s} u\rangle^m_{j^-}}{\Delta x}+
\frac{\langle p_\mathrm{s}u \rangle^m_{j^+}-\langle p_\mathrm{s} u\rangle^m_{j^-}}{\Delta x}\notag\\
-\frac{\rho^k_j+\rho^l_j}{4\rho^k_j \rho^l_j}\{(\rho u)^k_j+(\rho u)^l_j \}
\frac{\langle p_\mathrm{s} \rangle^m_{j^+}-\langle p_\mathrm{s} \rangle^m_{j^-}}{\Delta x},\label{eq:a.2}\\
DFS^{k,l,m}_{\mathrm{s},j}\equiv -\frac{\rho^k_j+\rho^l_j}{8\rho^k_j \rho^l_j}\{(\rho u)^k_j+(\rho u)^l_j \}
\frac{\langle A \rangle^m_{j^+}-\langle A \rangle^m_{j^-}}{\Delta x}+
\frac{\langle B_\mathrm{s} \rangle^m_{j^+}-\langle B_\mathrm{s} \rangle^m_{j^-}}{\Delta x},\label{eq:a.3}\\
ERR^{k,l,m}_{\mathrm{s},j}\equiv -\frac{(\rho^2 u^2)^k_j+(\rho^2 u^2)^l_j}{8\rho^k_j\rho^l_j}
\frac{\langle \rho u \rangle^m_{j^+}-\langle \rho u \rangle^m_{j^-}}{\Delta x}+
\frac{\rho^k_j+\rho^l_j}{8\rho^k_j \rho^l_j}\{(\rho u)^k_j+(\rho u)^l_j \}
\frac{\langle \rho u^2 \rangle^m_{j^+}-\langle \rho u^2 \rangle^m_{j^-}}{\Delta x}\notag\\
-\frac14\frac{\langle \rho u^3\rangle^m_{j^+}-\langle \rho u^3\rangle^m_{j^-}}{\Delta x},\label{eq:a.4}
\end{align}
where $ADV$, $DFS$, and $ERR$ represent the advection, artificial dissipation, and error components, respectively.

\subsection{Second-order RK method}\label{sec:a2}
Here, the proposed scheme is extended using the second-order RK method.
The conservation laws of mass, momentum, and energy are discretized as follows:
\begin{align}
\frac{\rho^{*}_j-\rho^n_j}{\Delta t}+\frac{\langle\rho u\rangle^n_{j^+}
-\langle\rho u\rangle^n_{j^-}}{\Delta x}=0,\label{eq:a.5}\\
\frac{(\rho u)^{*}_j-(\rho u)^n_j}{\Delta t}+
\frac{\langle\rho u^2+p\rangle^n_{j^+}-
\langle\rho u^2+p\rangle^n_{j^-}}{\Delta x}=
\frac{\langle A\rangle^n_{j^+}-\langle A\rangle^n_{j^-}}{\Delta x},\label{eq:a.6}\\
\frac{(\rho e+\frac12\rho u^2)^{*}_j-(\rho e+\frac12\rho u^2)^n_j}{\Delta t}
+\frac{\langle\rho eu+\frac12\rho u^3+p u\rangle^n_{j^+}-
\langle\rho eu+\frac12\rho u^3+pu\rangle^n_{j^-}}{\Delta x}=
\frac{\langle B\rangle^n_{j^+}-\langle B\rangle^n_{j^-}}{\Delta x},\label{eq:a.7}\\
\frac{\rho^{n+1}_j-\rho^n_j}{\Delta t}
+\frac{\langle\rho u\rangle^n_{j^+}-\langle\rho u\rangle^n_{j^-}}{2\Delta x}
+\frac{\langle\rho u\rangle^*_{j^+}-\langle\rho u\rangle^*_{j^-}}{2\Delta x}=0,\label{eq:a.8}\\
\frac{(\rho u)^{n+1}_j-(\rho u)^n_j}{\Delta t}
+\frac{\langle\rho u^2+p\rangle^n_{j^+}-\langle\rho u^2+p\rangle^n_{j^-}}{2\Delta x}
+\frac{\langle\rho u^2+p\rangle^*_{j^+}-\langle\rho u^2+p\rangle^*_{j^-}}{2\Delta x}\notag\\
=\frac{\langle A\rangle^n_{j^+}-\langle A\rangle^n_{j^-}}{2\Delta x}
+\frac{\langle A\rangle^*_{j^+}-\langle A\rangle^*_{j^-}}{2\Delta x},\label{eq:a.9}\\
\frac{(\rho e+\frac12\rho u^2)^{n+1}_j-(\rho e+\frac12\rho u^2)^n_j}{\Delta t}
+\frac{\langle\rho eu+\frac12\rho u^3+p u\rangle^n_{j^+}-\langle\rho eu+\frac12\rho u^3+pu\rangle^n_{j^-}}{2\Delta x}\notag\\
+\frac{\langle\rho eu+\frac12\rho u^3+p u\rangle^*_{j^+}-\langle\rho eu+\frac12\rho u^3+pu\rangle^*_{j^-}}{2\Delta x}
=\frac{\langle B\rangle^n_{j^+}-\langle B\rangle^n_{j^-}}{2\Delta x}
+\frac{\langle B\rangle^*_{j^+}-\langle B\rangle^*_{j^-}}{2\Delta x},\label{eq:a.10}
\end{align}
where ``$*$'' denotes the internal time-step of the second-order RK method.
Equations~(\ref{eq:a.5})--(\ref{eq:a.7}) and (\ref{eq:a.8})--(\ref{eq:a.10}) correspond to
the primary and secondary RK steps, respectively.
Obviously, the following energy equations are obtained from Eq.~(\ref{eq:a.5})--(\ref{eq:a.7}):
\begin{align}
ADV^{*,n,n}_{\mathrm{s},j}=DFS^{*,n,n}_{\mathrm{s},j}+ERR^{*,n,n}_{\mathrm{s},j}.\label{eq:a.11}
\end{align}
In addition, the spatial-difference terms of Eq.~(\ref{eq:a.8})--(\ref{eq:a.10}) can be interpreted as
arithmetic averages of ``$n$'' and ``$*$'' steps.
Thus, the energy equations in the second step are expressed as
\begin{align}
\frac{ADV^{n+1,n,n}_{\mathrm{s},j}+ADV^{n+1,n,*}_{\mathrm{s},j}}{2}=
\frac{DFS^{n+1,n,n}_{\mathrm{s},j}+DFS^{n+1,n,*}_{\mathrm{s},j}}{2}+
\frac{ERR^{n+1,n,n}_{\mathrm{s},j}+ERR^{n+1,n,*}_{\mathrm{s},j}}{2}.\label{eq:a.12}
\end{align}

\subsection{Third-order TVD RK method}
The implementation of the third-order TVD RK method is derived using a similar method to the second-order case.
The first step is expressed as
\begin{align}
ADV^{\dagger,n,n}_{\mathrm{s},j}=DFS^{\dagger,n,n}_{\mathrm{s},j}+ERR^{\dagger,n,n}_{\mathrm{s},j},\label{eq:a.13}
\end{align}
where ``$\dagger$'' indicates the primary internal time-step of the third-order TVD RK.
The second step is given as
\begin{align}
\frac{ADV^{\dagger\dagger,\dagger,\dagger}_{\mathrm{s},j}-3ADV^{\dagger\dagger,\dagger,n}_{\mathrm{s},j}}{4}=
\frac{DFS^{\dagger\dagger,\dagger,\dagger}_{\mathrm{s},j}-3DFS^{\dagger\dagger,\dagger,n}_{\mathrm{s},j}}{4}+
\frac{ERR^{\dagger\dagger,\dagger,\dagger}_{\mathrm{s},j}-3ERR^{\dagger\dagger,\dagger,n}_{\mathrm{s},j}}{4},\label{eq:a.14}
\end{align}
where ``$\dagger\dagger$'' indicates the secondary internal time-step of the third-order TVD RK.
The final step is obtained by weighting these internal steps, such that
\begin{align}
\frac{ADV^{n+1,n,n}_{\mathrm{s},j}+ADV^{n+1,n,\dagger}_{\mathrm{s},j}+4ADV^{n+1,n,\dagger\dagger}_{\mathrm{s},j}}{6}=\notag\\
\frac{DFS^{n+1,n,n}_{\mathrm{s},j}+DFS^{n+1,n,\dagger}_{\mathrm{s},j}+4DFS^{n+1,n,\dagger\dagger}_{\mathrm{s},j}}{6}+
\frac{ERR^{n+1,n,n}_{\mathrm{s},j}+ERR^{n+1,n,\dagger}_{\mathrm{s},j}+4ERR^{n+1,n,\dagger\dagger}_{\mathrm{s},j}}{6}.\label{eq:a.15}
\end{align}

\section{Multidimensional description} \label{sec:b}
Multidimensional scheme is required for the numerical simulations of ICF implosions
because hydrodynamic instabilities such as Rayleigh--Taylor instability \cite{Takabe1985,Betti1998} are
one of the fundamental physics to determine the fusion gain.
Here, we introduce a multidimensional description of the proposed approach.
Before the derivation of multidimensional scheme,
we define two vector differential operators in the discrete form:
\begin{align}
\mathrm{Grad\ } \phi|^n_{i,j,k}=\begin{bmatrix}
\dfrac{\langle \phi \rangle^n_{i+\frac12,j,k}-\langle \phi \rangle^n_{i-\frac12,j,k}}{\Delta x} \\
\dfrac{\langle \phi \rangle^n_{i,j+\frac12,k}-\langle \phi \rangle^n_{i,j-\frac12,k}}{\Delta y} \\
\dfrac{\langle \phi \rangle^n_{i,j,k+\frac12}-\langle \phi \rangle^n_{i,j,k-\frac12}}{\Delta z}
\end{bmatrix},\label{eq:b.1}\\
\mathrm{Div\ } \mathbf{\Psi}|^n_{i,j,k}=
\dfrac{\langle \psi_{x} \rangle^n_{i+\frac12,j,k}-\langle \psi_{x}\rangle^n_{i-\frac12,j,k}}{\Delta x}+
\dfrac{\langle \psi_{y} \rangle^n_{i,j+\frac12,k}-\langle \psi_{y}\rangle^n_{i,j-\frac12,k}}{\Delta y}+
\dfrac{\langle \psi_{z} \rangle^n_{i,j,k+\frac12}-\langle \psi_{z}\rangle^n_{i,j,k-\frac12}}{\Delta z}
\end{align}
where $\phi$ is an arbitrary scalar function, $\mathbf{\Psi}=^\mathrm{T}[\psi_x,\psi_y,\psi_z]$
is an arbitrary vector function, and $(i, j, k)$ are the spatial indices over $(x, y, z)$ directions, respectively.
The three-dimensional Euler equation is discretized by using these operators as follows:
\begin{align}
\frac{\rho^{n+1}_{i,j,k}-\rho^n_{i,j,k}}{\Delta t}+\mathrm{Div\ }(\rho \mathbf{u})|^n_{i,j,k}=0,\label{eq:b.3}\\
\frac{(\rho \mathbf{u})^{n+1}_{i,j,k}-(\rho \mathbf{u})^n_{i,j,k}}{\Delta t}+\mathrm{Div\ }(\rho \mathbf{uu})|^n_{i,j,k}
+\mathrm{Grad\ } p|^n_{i,j,k} =\mathbf{0},\label{eq:b.4}\\
\frac{(\rho e+\frac12\rho |\mathbf{u}|^2)^{n+1}_{i,j,k}-(\rho e+\frac12\rho |\mathbf{u}|^2)^n_{i,j,k}}{\Delta t}
+\mathrm{Div\ }\left.\left( \rho e \mathbf{u} + \frac12\rho|\mathbf{u}|^2\mathbf{u} + p\mathbf{u}\right)\right|^n_{i,j,k}=
0,\label{eq:b.5}
\end{align}
Note that the key of our approach in Sec.~\ref{sec:2.2} is the expansion of the time derivative.
Hence, the three-dimensional scheme is also obtained by the same way.
The discretized equations corresponding to Eqs.~(\ref{eq:1.3}) and (\ref{eq:1.4}) are
\begin{align}
\frac{(\rho e_\mathrm{s})^{n+1}_{i,j,k}-(\rho e_\mathrm{s})^n_{i,j,k}}{\Delta t}+
\mathrm{Div\ }(\rho e_\mathrm{s} \mathbf{u}+p_\mathrm{s} \mathbf{u})|^n_{i,j,k}-
\frac{\rho^{n+1}_{i,j,k}+\rho^n_{i,j,k}}{4\rho^{n+1}_{i,j,k}\rho^n_{i,j,k}}
\{(\rho \mathbf{u})^{n+1}_{i,j,k}+(\rho \mathbf{u})^n_{i,j,k}\}\cdot
\left(\mathrm{Grad\ }p_\mathrm{s}|^n_{i,j,k}\right)=\notag\\
-\frac{(|\rho \mathbf{u}|^2)^{n+1}_{i,j,k}+(|\rho \mathbf{u}|^2)^n_{i,j,k}}{8\rho^{n+1}_{i,j,k}\rho^n_{i,j,k}}
\mathrm{Div\ }(\rho \mathbf{u})|^n_{i,j,k}
+\frac{\rho^{n+1}_{i,j,k}+\rho^n_{i,j,k}}{8\rho^{n+1}_{i,j,k}\rho^n_{i,j,k}}
\{(\rho \mathbf{u})^{n+1}_{i,j,k}+(\rho \mathbf{u})^n_{i,j,k}\}\cdot\mathrm{Div\ }(\rho \mathbf{uu})|^n_{i,j,k}\notag\\
-\mathrm{Div\ }\left.\left(\frac14\rho|\mathbf{u}|^2\mathbf{u}\right)\right|^n_{i,j,k}.
\end{align}
Extension to the shock capturing scheme is self-evident.


\end{document}